\documentclass{procmult}
\usepackage{graphicx}

\begin{document}

\newcommand\be{\begin{equation}}
\newcommand\ee{\end{equation}}
\newcommand\bea{\begin{eqnarray}}
\newcommand\eea{\end{eqnarray}}
\newcommand\bseq{\begin{subequations}} 
\newcommand\eseq{\end{subequations}}
\newcommand\bcas{\begin{cases}}
\newcommand\ecas{\end{cases}}
\newcommand{\p}{\partial}
\newcommand{\f}{\frac}
\renewcommand{\a}{\alpha}
\renewcommand{\b}{\beta}
\renewcommand{\k}{\kappa}

\title{Semi-classical isotropization of the Mixmaster Universe}
\author{Marco Valerio Battisti$^\dag$, Riccardo Belvedere$^\ddag$ \and Giovanni Montani$^\S$}
\institute{\small{
$^\dag$ Centre de Physique Th\'eorique de Luminy, Universit\'e de la M\'editerran\'ee, F-13288 Marseille FR, and ``Sapienza'' Universit\`a di Roma, Dipartimento di Fisica and ICRA,\\ P.le A. Moro 5, 00185 Rome IT, battisti@icra.it \\
$^\ddag$ ``Sapienza'' Universit\`a di Roma, Dipartimento di Fisica and ICRA, P.le A. Moro 5, 00185 Rome IT, riccardo.belvedere@icra.it \\
$^\S$ ICRA, ICRANet, ENEA and ``Sapienza'' Universit\`a di Roma, Dipartimento di Fisica, P.le A. Moro 5, 00185 Rome IT, montani@icra.it}
}
\maketitle

\abstract{A semi-classical mechanism which leads to the isotropization of the Mixmaster Universe is developed. A wave function of this Universe, which has a meaningful probabilistic interpretation, is constructed and it describes the evolution of the anisotropies of the Universe with respect to the isotropic scale factor, which plays the external observer-like role. We show that, once large volume regions are investigated, the closed Friedmann-Robertson-Walker configuration is deeply privileged.}

\bigskip

Quantum cosmology denotes the application of the quantum theory to the entire Universe \cite{QC}. It can be then viewed as a natural arena to investigate as part of a more general drive to understood quantum gravity. In canonical quantum gravity, the quantum state $\Psi$ of the system is generally represented by a wave functional describing the dynamics of the three metric $h_{ij}$ as well as matter fields $\phi$, i.e.  $\Psi=\Psi[h_{ij}(x),\phi(x)]$. This state is defined on the the Wheeler superspace and, since of the diffeomorphisms invariance of GR, it has no explicit dependence on time. In particular, it (formally) satisfies the Dirac quantum implementation of the first-class constraints of GR.

A complete quantum theory of gravity is not yet available and therefore this problem is not properly defined. To overcame such a feature the fields are usually restricted (by hand) to a finite dimensional subspace of the superspace, i.e. we deal with the {\it minisuperspace representation}. Quantum cosmology is explicitly defined as the minisuperspace quantization of homogeneous (finite degrees of freedom) cosmological models. From the peculiarity of the system-Universe, fundamental interpreting difficulties of the wave function of the Universe $\Psi$ arise. The question about the interpretation (i.e. extracting physical statements) of quantum cosmology clearly appears as soon as the differences with respect to ordinary quantum mechanics are addressed \cite{Vil,pqc}. The standard interpretation of quantum mechanics (the Copenhagen one) involves the following basic assumptions. (i) There exist an external observer to the quantum system, i.e. the model under investigation is not genuinely closed. (ii) Predictions are probabilistic in nature and performed by a measurement of an external agency. (iii) Time plays a central and peculiar role. By contrast quantum cosmology is defined by the following features. (i) The analyzed model is the Universe as whole, i.e. it is closed without external observers. (ii) No external measurement crutch is available and an internal one can not plays the observer-like role since of the extreme conditions a very early Universe is subjected on. (iii) The time coordinate is not an observable in GR and at quantum level it is known as the problem of time.  

The most developed idea to solve these features relies in accepting that a meaningfully interpretation of the wave function of the Universe can be only formulated at semi-classical level. More precisely, it is only possible to quantum-mechanically interpret a small subsystem of the entire Universe, i.e. in the domain where at least some of the minisuperspace variables are semi-classical in the sense of the Wentzel-Kramers-Brillouin (WKB) approximation. 

In this work such a scheme is implemented to the most general homogeneous cosmological model, i.e. the Mixmaster Universe. The relevance of this model relies on the fact that a {\it generic} solution of the Einstein equations toward the cosmological singularity is formulated by a collection of causal independent Bianchi IX horizons \cite{BKL}. In particular, a wave function of the Universe which has a clear probabilistic interpretation when the isotropic scale factor $a$ of the Universe is regarded as semi-classical is obtained. It describes the quantum evolution of the Mixmaster anisotropies and its dynamics is traced with respect to $a$, which can be regarded as a semi-classical variable as soon as the Universe expands enough. The main result is that the wave function of the Universe is spread over all values of anisotropy near the cosmological singularity but, when the radius of the Universe grows, it is asymptotically peaked around the isotropic configuration. The closed Friedmann-Robertson-Walker (FRW) cosmological model is thus the naturally privileged state far enough from the classical singularity. A semi-classical isotropization mechanism for the Mixmaster Universe is then predicted.

The dynamics of the homogeneous cosmological models (the Bianchi Universes) is summarized, in the canonical formalism, by the scalar constraint (for reviews see \cite{rev})
\be\label{scacon} 
\mathcal H=\kappa\left[-\f{p_a^2}a+\f1{a^3}\left(p_+^2+p_-^2\right)\right]+\f{a}{4\kappa}V(\beta_\pm)+U(a)=0,
\ee  
where the potential term $V(\beta_\pm)$ accounts for the spatial curvature of the model. Here $\kappa=8\pi G$, the variable $a=a(t)$ describes the isotropic expansion of the Universe and its shape changes (anisotropies) are associated to $\beta_\pm=\beta_\pm(t)$. The phase space of this model is thus six dimensional and the cosmological singularity appears for $a\rightarrow0$. In the Universe dynamics we have assumed the matter terms to be negligible with respect to the cosmological constant $\Lambda$, i.e. the isotropic potential $U(a)$ reads $U(a)=-a/4\kappa+\Lambda a^3/\kappa$. As matter of fact, far enough from the singularity, the cosmological constant term dominates on the other ordinary matter fields and such a contribution is necessary in order to the inflationary scenario takes place \cite{KM,Kolb}. 

As we said, a correct definition of probability in quantum cosmology can be formulated by distinguishing between semi-classical and quantum variables \cite{Vil}. More precisely, the variables which satisfy the Hamilton-Jacobi equation are regarded as semi-classical and is assumed that the quantum variables do not affect the dynamics generated by the semi-classical ones. In this respect we claim that the quantum variables describe a small subsystem of the Universe and is then natural to regard the isotropic expansion variable $a$ as the semi-classical one while considering the anisotropy coordinates $\beta_\pm$ (the two physical degrees of freedom of the Universe) as the purely quantum variables. We are thus requiring ab initio that the radius of the Universe is of different nature with respect to the anisotropies. To implement such a picture, the wave function of the Universe $\Psi=\Psi(a,\beta_\pm)$ is assumed to be \cite{Vil} 
\be\label{wvil}
\Psi=\Psi_0\chi=A(a)e^{iS(a)}\chi(a,\beta_\pm).
\ee 
This wave function is WKB-like in the $a$ coordinate and the additional function $\chi$ depends on the quantum variables $\beta_\pm$ and only parametrically, in the sense of the Born-Oppenheimer approximation, on $a$.

The canonical quantization of this model is achieved by the use of the Dirac prescription for quantizing constrained systems \cite{HT}, i.e. imposing that the physical states are those annihilated by the self-adjoint operator $\hat{\mathcal H}$ corresponding to the classical counterpart (\ref{scacon}). Considering (\ref{wvil}), we obtain from the quantum operator version of (\ref{scacon}) the Hamilton-Jacobi equation for $S$ and the continuity equation for the amplitude $A$
\be\label{hj}
-\kappa A\left(S'\right)^2+aUA+\mathcal V_q=0, \qquad \f1A\left(A^2S'\right)'=0,
\ee
respectively. Here the prime denotes differentiation with respect to the scale factor $a$ and $\mathcal V_q=\kappa A''$ is the so-called quantum potential, which in this model is negligible far from the classical singularity even if the $\hbar\rightarrow0$ limit is not taken into account (see below). As usual $S(a)$ defines a congruence of classical trajectories. The new equation we find is a Schr\"odinger-like one describing the evolution of the proper quantum state $\chi$. Neglecting higher order correction terms in $\hbar$, it reads
\be\label{eveq}
-2ia^2S'\p_a\chi=\hat H_q\chi, \qquad H_q=p_+^2+p_-^2+\f{a^4}{4\kappa^2}V(\beta_\pm).
\ee
Such an equation is in agreement with the assumption that the anisotropies describe a quantum subsystem of the whole Universe, i.e. that the wave function $\chi$ depends only on $\beta_\pm$ (in the Born-Oppenheimer sense). As matter of fact, the smallness of such a quantum subsystem can be formulated requiring that its Hamiltonian $H_q$ is of order $\mathcal O(\epsilon^{-1})$, where $\epsilon$ is a small parameter proportional to $\hbar$. Since the action of the semi-classical Hamiltonian operator $\hat H_0=a^2\p_a^2+a^3U/\kappa$ on the wave function $\Psi$ is of order $\mathcal O(\epsilon^{-2})$, the idea that the anisotropies do not influence the isotropic expansion of the Universe can be formulated as $\hat H_q\Psi/\hat H_0\Psi=\mathcal O(\epsilon)$. Such a requirement is physically reasonable since, the semi-classical proprieties of the Universe as well as the smallness of the quantum subsystem, are both related to the fact that the Universe is large enough \cite{Vil}.  

A purely Schr\"odinger equation for the wave function $\chi$ is obtained taking into account the tangent vector to the classical path. Using $p_a=S'$, the equations of motion (\ref{hj}) and considering the time gauge $da/dt=1$, is possible to define the new time variable $\tau$ such that $d\tau=(N\kappa/a^3)da$. In the asymptotic interesting region ($a\gg l_\Lambda\equiv1/\sqrt\Lambda$) the evolution equation (\ref{eveq}) rewrites as
\be\label{scheq}
i\p_\tau\chi=\left(-\Delta_\beta+\f{a^4}{4\kappa^2}V(\beta_\pm)\right)\chi,
\ee
where $\tau=(\kappa/12\sqrt\Lambda)a^{-3}+\mathcal O(a^{-5})$. This is the Schr\"odinger equation for the wave function $\chi$ describing the quantum variables $\beta_\pm$. The wave function (\ref{wvil}) defines a probability distribution $\rho(a,\beta_\pm)$ which appears to be $\rho(a,\beta_\pm)=\rho_0(a)\rho_\chi(a,\beta_\pm)$, where $\rho_0(a)$ is the classical probability distribution for the semi-classical variable $a$ and $\rho_\chi=|\chi|^2$ denotes the probability distribution for the quantum variables $\beta_\pm$ on the classical trajectories (\ref{hj}) where the wave function $\chi$ can be normalized. 

In order to enforce the idea that the anisotropies can be considered as the only quantum degrees of freedom of the Universe, we consider the quasi-isotropic regime $|\beta_\pm|\ll1$. Moreover, since we are interested at the link between the isotropic and anisotropic dynamics, the Universe has to be get through to such a quasi-isotropic era. In this regime, the potential term reads $V(\beta_\pm)=8(\beta_+^2+\beta_-^2)+\mathcal O(\beta^3)$ and the Schr\"odinger equation (\ref{scheq}) can be then written as 
\be\label{osceq}
i\p_\tau\chi=\f12\left(-\Delta_\beta+\omega^2(\tau)(\beta_+^2+\beta_-^2)\right)\chi,
\ee
where $\omega^2(\tau)=C/\tau^{4/3}$ and $C$ being a constant given by $2C=1/6^{4/3}(\kappa\Lambda)^{2/3}$. In other words, we are dealing with a time-dependent bi-dimensional harmonic oscillator with frequency $\omega(\tau)$. The quantum theory of an harmonic oscillator with time-dependent frequency is well known \cite{harosc} and the solution of the Schr\"odinger equation (\ref{osceq}) can be analytically obtained. Through the introduction of the generalized invariant state, whose eigenstates are connected with those of a time-independent harmonic oscillator, and via an unitary transformation, the wave function $\chi_n=\chi_+\chi_-$ reads
\be\label{chi}
\chi_\pm=\chi_n(\beta_\pm,\tau)=A\f{e^{i\alpha_n(\tau)}}{\sqrt\rho}h_n(\beta_\pm/\rho)\exp\left[\f i2\left(\dot\rho\rho^{-1}+i\rho^{-2}\right)\beta_\pm^2\right].
\ee 
In this formula $A$ denotes the normalization constant, $h_n$ are the usual Hermite polynomial of order $n$ and $\rho(\tau)$ and the phase $\alpha(\tau)$ are respectively given by
\be\label{rho}
\rho=\sqrt{\f\tau{\sqrt C}\left(1+\f{\tau^{-2/3}}{9C}\right)}, \qquad \alpha_n=-\left(n+\f12\right)\int\f{d\tau}{\rho^2(\tau)}.
\ee
It is immediate to verify that, when $\omega(\tau)\rightarrow\omega_0$ and $\rho(\tau)\rightarrow\rho_0=1/\sqrt{\omega_0}$ (namely $\alpha(\tau)\rightarrow-\omega_0(n+1/2)\tau$), the solution of a time-independent harmonic oscillator is recovered.

Let us now investigate the probability density to find the quantum subsystem of the Universe at a given state. The anisotropies appear to be probabilistically suppressed as soon as the Universe expands enough far from the cosmological singularity (which we remember appears for $a\rightarrow0$ or $\tau\rightarrow\infty$). Such a feature can be immediately observed from the behavior of the squared modulus of the wave function (\ref{chi}) which is given by
\be\label{prob}
|\chi_n|^2\sim\f1{\rho^2}|h_{n_+}(\beta_+/\rho)|^2|h_{n_-}(\beta_-/\rho)|^2 e^{-\beta^2/\rho^2},
\ee  
where $\beta^2=\beta^2_++\beta^2_-$ and with $\sim$ we omit the normalization constant. This probability density is still time-dependent through $\rho=\rho(\tau)$ since the evolution of the wave function $\chi$ is not traced by an unitary time operator. As we can see from (\ref{prob}), when a large enough isotropic cosmological region is considered (namely when the limit $a\rightarrow\infty$ or $\tau\rightarrow0$ is taken into account), the probability density to find the Universe is sharply peaked at the isotropic configuration, i.e. for $|\beta_\pm|\simeq0$. In this limit (which corresponds to $\rho\rightarrow0$) the probability density $|\chi_{n=0}|^2$ of the ground state ($n=n_++n_-=0$) is given by  $|\chi_{n=0}|^2\stackrel{\tau\rightarrow0}\longrightarrow\delta(\beta,0)$, thus is proportional to the Dirac $\delta$-distribution centered on $\beta=0$ (see Fig. 1). 
\begin{figure}
\centering
\includegraphics[height=1.8in]{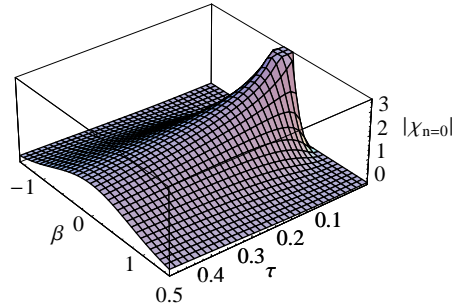}
\caption{The ground state of the wave function $\chi(\beta_\pm,\tau)$ far from the cosmological singularity, i.e. in the $\tau\rightarrow0$ limit. In the plot we take $C=1$.} 
\end{figure} 

Summarizing, when the Universe moves away from the cosmological singularity, the probability density to find it is asymptotically peaked around the closed FRW configuration. Near the initial singularity all values of anisotropy $\beta$ are almost equally favored from a probabilistic point of view. On the other hand, as the radius of the Universe grows, the isotropic state become the most probable one. (For other similar approaches see \cite{qiso}.)

It is worth noting that the key feature of such a result relies on the fact that the isotropic scalar factor $a$ was considered as an intrensically different variable with respect to the anisotropies. It was treated as a semi-classical variable while only the two physical degrees of freedom of the Universe ($\beta_\pm$) were described as real quantum coordinates. This way, a positive semidefinite probability density for the wave function of the quantum subsystem of the Universe can be constructed and a clear interpretation of it considered. The validity of such an assumption can be verified from the analysis of the Hamilton-Jacobi equations (\ref{hj}). In particular, the WKB function $\Psi_0=\exp(iS+\ln A)$ approaches the quasi-classical limit $e^{iS}$ as soon as $a\gg l_\Lambda$ ($l_\Lambda$ being the inflation characteristic length \cite{Kolb}). To corroborate the model, we have studied the classical limit too. Splitting the $S$-function in two terms, respectively for the time variable $a$ and for the anisotropies $\beta_\pm$, we achieve, if $a\gg l_\Lambda$, an analogous behavior of the anisotropies, i.e. them go to reduce themselves once one moves away from the cosmological singularity.

\bigskip

{\it Acknowledgments.} M. V. B. thanks ''Fondazione Angelo Della Riccia'' for financial support.

\end{document}